\documentclass[12pt]{article}
\usepackage{sectsty}
\usepackage{mathrsfs}
\allsectionsfont{\sffamily}
\newcommand{\5}{$_5$}

\usepackage[nosort]{cite}

\usepackage[pdftex]{graphicx}
\usepackage[font={scriptsize,singlespacing}]{caption}
\usepackage{epstopdf}

\usepackage{amsmath}
\usepackage{amssymb}
\usepackage{subfigure}
\usepackage{hyperref}
\usepackage{url}
\usepackage{xcolor}
\usepackage{color}
\definecolor{amaranth}{rgb}{0.9, 0.17, 0.31}
\definecolor{purple(munsell)}{rgb}{0.62, 0.0, 0.77}
\definecolor{americanrose}{rgb}{1.0, 0.01, 0.24}
\definecolor{palatinateblue}{rgb}{0.15, 0.23, 0.89}
\definecolor{royalblue(web)}{rgb}{0.25, 0.41, 0.88}
\definecolor{hanpurple}{rgb}{0.32, 0.09, 0.98}
\definecolor{beaublue}{rgb}{0.74, 0.83, 0.9}
\definecolor{carminered}{rgb}{1.0, 0.0, 0.22}
\definecolor{brightpink}{rgb}{1.0, 0.0, 0.5}
\hypersetup{ linktoc=all,
    colorlinks, linkcolor={palatinateblue},
    citecolor={brightpink}, urlcolor={purple(munsell)}
}

\def\sideremark#1{\ifvmode\leavevmode\fi\vadjust{\vbox to0pt{\vss
 \hbox to 0pt{\hskip\hsize\hskip1em
 \vbox{\hsize2cm\tiny\raggedright\pretolerance10000
 \noindent #1\hfill}\hss}\vbox to8pt{\vfil}\vss}}}%
\newcommand{\bo}{\raise-1mm\hbox{\Large$\Box$}}

\newcommand{\f}[2]{\frac{#1}{#2}}

\newcommand{\la}{\langle}
\newcommand{\ra}{\rangle}
\newcommand{\w}{\omega}
\newcommand{\kp}{\kappa}
\newcommand{\be}{\begin{equation}}
\newcommand{\ee}{\end{equation}}
\newcommand{\bea}{\begin{eqnarray}}
\newcommand{\eea}{\end{eqnarray}}

\begin{document}
\thispagestyle{empty}
\begin{center}

\null \vskip-1truecm \vskip2truecm

{\Large{\bf \textsf{Signatures of Energy Flux in Particle Production:}}}
\vskip0.1truecm
{\Large{\bf \textsf{A Black Hole Birth Cry and Death Gasp\\}}}

\vskip1truecm
\textbf{\textsf{Michael R.R. Good}}\\
{\footnotesize\textsf{Department of Physics, Nazarbayev University, \\53 Kabanbay Batyr Ave., 
Astana, Republic of Kazakhstan}\\
{\tt Email: michael.good@nu.edu.kz}}\\

\vskip0.4truecm
\textbf{\textsf{Yen Chin Ong}}\\
{\footnotesize \textsf{Nordic Institute for Theoretical Physics, \\ KTH Royal Institute of Technology \& Stockholm University, \\ Roslagstullsbacken 23,
SE-106 91 Stockholm, Sweden}\\
{\tt Email: yenchin.ong@nordita.org}}\\

\end{center}
\vskip1truecm \centerline{\textsf{ABSTRACT}} \baselineskip=15pt

\medskip
It is recently argued that if the Hawking radiation process is unitary, then a black hole's mass cannot be monotonically decreasing. We examine the time dependent particle count and negative energy flux in the non-trivial conformal vacuum via the moving mirror approach. A new, exactly unitary solution is presented which emits a characteristic above-thermal positive energy burst, a thermal plateau, and negative energy flux.  It is found that the characteristic positive energy flare and thermal plateau is observed in the particle outflow.  However, the results of time dependent particle production show no overt indication of negative energy flux. Therefore, a black hole's birth cry is detectable by asymptotic observers via particle count, whereas its death gasp is not. 


\vskip0.4truecm
\hrule

\section{The Puzzle of the Black Hole's Last Gasp}

The final fate of a black hole under Hawking evaporation \cite{Hawking2} has long been a puzzle. The simplest model of Hawking evaporation, assuming thermal emission, is that the process (for an asymptotically flat Schwarzschild black hole, in the units $G=c=\hbar=k_B=1$) is governed by the black body equation (up to a greybody factor),
\begin{equation}\label{naive}
\frac{dM}{dt} = -AT^4 \sim -\frac{1}{M^2},
\end{equation}
where $A \sim M^2$ denotes the horizon area and $T \sim 1/M$ the Hawking temperature of the black hole. 
In particular, if this equation holds for all time, then it means that the black hole mass (for an isolated black hole) is a monotonically decreasing function of time. However, we know that if the Hawking process is unitary, then at some point about when the black hole loses half of its initial Bekenstein-Hawking entropy, information will start to ``leak out of the black hole''. This feature is captured by the turn-over of the Page curve \cite{page1, page1b, page2}. This does not mean that the thermal spectrum and Eq.(\ref{naive}) will necessarily be modified at late times, since the thermal spectrum does not necessitate zero-correlation between the radiation \cite{sabine, 1409.7754}. Nevertheless, taking into account other effects, it has been argued that Hawking radiation is only approximately Planck-shaped over an explicitly bounded range of frequencies \cite{1409.7754}.

Despite all these subtleties, the usual thought is that Hawking radiation can only decrease the mass of the black hole.
In particular, Eq.(\ref{naive}) certainly implies monotonic mass loss, and assuming this holds even at late times, the black hole has finite life time of the order $M^3$.
However, it was recently shown by Bianchi and Smerlak \cite{Bianchi:2014vea, Bianchi:2014qua} that, surprisingly --- under the reasonable assumption that unitarity is preserved --- the mass loss of a black hole is not even monotonically decreasing. That is, the mass of an evaporating black hole can \emph{temporarily increase}. The physical picture is as follows: in the usual scenario of Hawking evaporation, a positive flux of energy is received at infinity, corresponding to the mass loss of the black hole. However, a \emph{negative} flux of energy is also eventually received at infinity\footnote{See also Figure 12 in \cite{HMF}, in which a moving mirror model was used to investigate the black hole firewall issue \cite{amps, apologia, sam}.}.  The puzzle is the following: what exactly causes the black hole to ``swell up'' --- the black hole's last gasp \cite{Bianchi:2014vea}? We have no good physical intuition for such a process, however in this work we show that the particle emission process gives no indication that the black hole is taking its last gasp.  

In other words, \emph{the negative energy flux (NEF) is not characterized by any hint in the particle emission.} Therefore, an asymptotic observer who is equipped with a particle detector and only has access to the particle emission at null infinity, does not detect any unusual particle emission, which might otherwise have indicated that the black hole is undergoing some unusual process. 
On the other hand, we show that the thermal flux, and even the above-thermal burst (which occurs in our model before the thermal flux) --- the black hole's birth cry, is detectable via particle count. 

Due to the difficulty with modeling Hawking evaporation, we shall work in 1+1 dimensions in natural units, $\hbar=c=1$. In (1+1)-dimensional spacetime, Newton's constant $G$ has units $L^1 M^{-1} T^{-2}$ to conform to the action principle, rather than the usual (3+1)-dimensional units of $L^3 M^{-1} T^{-2}$. (This $T$ is of course not to be confused with temperature.) We leave $G$ unassigned as we will work in flat spacetime.  No unique Planck time exists because there is no combination of $x,y,z$ for the dimensional analysis\footnote{In $D$-spacetime dimensions, the Planck time is $T_\text{Pl}=\left(\frac{G\hbar}{c^{D+1}}\right)^{\frac{1}{D-2}}$. Note that the $D=2$ case is special.} $[G^x\hbar^y c^z] =  \left(\frac{L^1}{M T^2}\right)^x\left(\frac{L^2 M}{T}\right)^y  \left(\frac{L}{T}\right)^z = T^1 $. A benefit to this unscaled time, is that we are allowed to choose time-sampling to be as small as we like without running up against Planck scale values.  We also employ the moving mirror method to model the evaporating black hole spacetime. A moving mirror refers to a free field theory in flat space with the boundary condition that the field
vanishes on the worldline of a reflecting point (the ``mirror'') \cite{weinstein}.  An accelerating mirror produces particles via a process called the \emph{dynamical Casimir effect}. Recently, 
the first experimental evidence for observation of the dynamical Casimir effect was reported \cite{Wilson:2011}; see also \cite{Rego:2014wta}. 

We are interested in two questions: ``Does particle production contain features of the energy flux?'';  and if so, ``Does the particle production contain hints of negative energy flux?''  




\section{An Exact Unitary Solution with Notable Attributes}
We first set up the background in Section (\ref{sec:background}) to explore a unitary solution that has features that can be probed both in terms of the particle production and the energy flux.  We then present the solution and check its consistency in Section (\ref{sec:solution}).  There are three notable attributes: a positive energy burst, a thermal plateau, and a negative energy flux.

\subsection{Some Background: Introducing the Moving Mirror} \label{sec:background} 
\begin{figure}[!h]
\centering
\mbox{\subfigure{\includegraphics[width=3.0in]{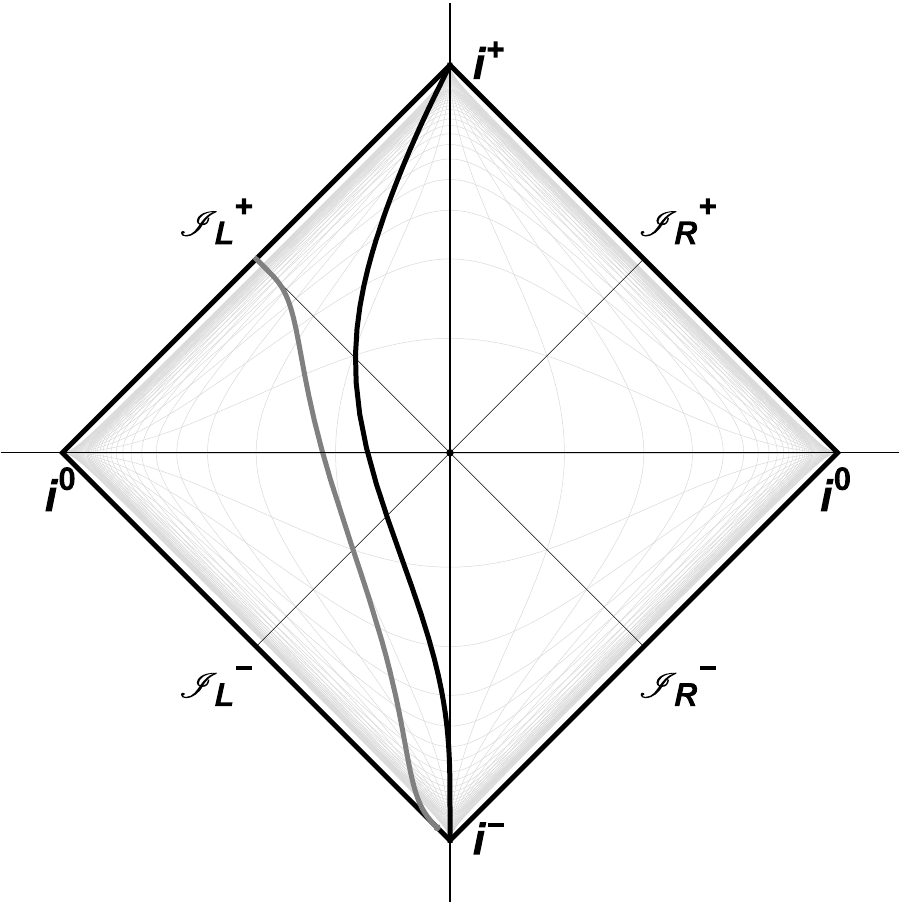}}\quad
\subfigure{\rotatebox{90}{\includegraphics[width=3.0in]{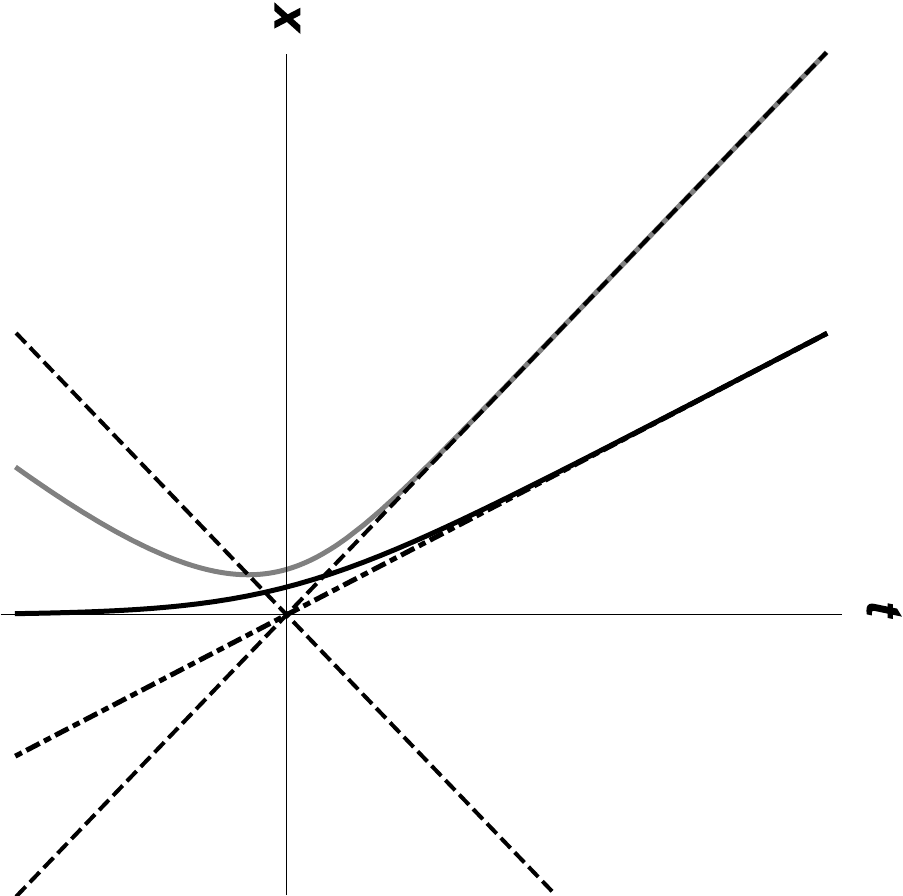} }}}
\caption{\textbf{Left:} In this Penrose diagram, the solid black line indicates the asymptotically inertial trajectory, exhibiting no horizon.  The grey line is the eternally thermal moving mirror. \textbf{Right:} The asymptotically inertial trajectory with a final coasting speed of half the speed of light displayed in the usual spacetime diagram.  The dashed lines represent the light cone, and the dotted-dashed line shows the asymptote of the mirror trajectory.  The trajectory example here is the same as in the conformal diagram.  For comparison, the grey line indicates the enternally thermal moving mirror\cite{Carlitz:1986nh} \cite{Good:2013lca}, which contains a horizon coinciding with the light cone. \label{fig:Penrose}}
\end{figure}

The moving mirror model is the most simple example of the dynamical Casimir effect \cite{Davies:1976hi} \cite{Davies:1977yv}.  It offers a way to understand black hole evaporation.  Consider a moving mirror as an external boundary condition in 1+1 dimensions, which \textit{does not} accelerate forever.  Consider it starting and ending at time-like past infinity $i^-$, and time-like future infinity $i^+$, respectively, possessing  asymptotically zero acceleration in both the far past and far future.  As such, an asymptotically inertial mirror will contain no horizons or pathological acceleration singularities, as seen by the solid black line in Figures (\ref{fig:Penrose}). The system, as we will see, is unitary.  

As a simple theoretical manifestation of black hole evaporation, the moving mirror model, has in practice, been very hard to solve for exact trajectories where the Bogolubov transformation may be evaluated, even in the (1+1)-dimensional case.  Very few soluble cases have been found \cite{Davies:1977yv, Carlitz:1986nh} and unitary cases are rare \cite{Walker:1982, Good:2013lca}.

The field is the massless scalar of the Klein-Gordon equation $\Box \psi = 0$, where the moving mirror is imposed at the $z(t)$ position such that the field is zero, $\psi|_z =0$.  The field has modes that can be used to expand it:
\be \psi = \int_0^\infty d\w'\; \left[a_{\w'} \phi_{\w'} + a^\dagger_{\w'}\phi^*_{\w'}\right] = \int_0^\infty d\w\; \left[b_\w \eta_\w + b^{\dagger}_\w\eta^{*}_\w\right]. \ee
The modes, being orthonormal and complete, take the null coordinate form,
\be 
\phi_{\w'} = (4\pi \w')^{-1/2} [e^{-i\w' v } - e^{-i\w' p(u)} ],
\ee
\be
\eta_{\w}  = (4\pi \w)^{-1/2} [e^{-i\w f(v)} - e^{-i\w u} ],
\ee
where the functions $p(u)$ and $f(v)$ are ray-tracing functions related to the trajectory of the mirror $z(t)$, see \cite{Good:2013lca}. Expanded modes in terms of the modes themselves, 
\be \label{phichi} \phi_{\w'} = \int_0^\infty d\w\; \left[\alpha_{\w'\w} \eta_{\w} + \beta_{\w'\w}\eta^*_{\w}\right], \ee
\be \label{chiphi} \eta_{\w} = \int_0^\infty d\w'\; \left[\alpha^*_{\w'\w} \phi_{\w'} - \beta_{\w'\w}\phi^*_{\w'}\right], \ee
is possible from the use of Bogolubov coefficients,  
\be \alpha_{\w'\w} = (\phi_{\w'},\eta_{\w}), \qquad \beta_{\w'\w} = -(\phi_{\w'},\eta^*_{\w'}), \ee
defined by the flat-space scalar product, expressed in null coodinates, 
\be (\phi_{\w'} , \eta_{\w} ) := i \int_{-\infty}^{\infty} du\; \phi_{\w'}^* \stackrel{\leftrightarrow}{\partial_u} \eta_{\w} + i \int_{-\infty}^{\infty} dv\; \phi_{\w'}^* \stackrel{\leftrightarrow}{\partial_v} \eta_{\w}. \ee
The Bogolubov coefficients $\alpha_{\w'\w}$ and $\beta_{\w'\w}$ also give the operators $a_{\w'}$ and $a_{\w'}^\dagger$ in terms of the operators $b_{\w}$ and $b^{\dagger}_{\w}$:
\be \label{aforBog} a_{\w'} =  \int d\w \left[\alpha^*_{\w'\w} b_{\w} - \beta^{*}_{\w'\w} b^{\dagger}_{\w}\right], \ee
\be \label{bforBog} b_{\w} =  \int d\w' \left[\alpha_{\w'\w} a_{\w'} + \beta^{*}_{\w'\w} a^{\dagger}_{\w'}\right]. \ee
%
The orthonormality of the modes are expressed via,
\be \label{phinormal}(\phi_{\w},\phi_{\w'}) = -(\phi^*_{\w},\phi^*_{\w'}) = \delta(\w-\w') ,\; (\phi_{\w},\phi^*_{\w'}) = 0, \ee
\be \label{chinormal}(\eta_{\w},\eta_{\w'}) = -(\eta^*_{\w},\eta^*_{\w'}) = \delta(\w-\w') ,\; (\eta_{\w},\eta^*_{\w'}) = 0. \ee
In the unitary case, the von Neumann entanglement entropy is found from Bianchi-Smerlak's formula\footnote{Here we set the constant $c$ (the central charge for a conformal field theory) in \cite{Bianchi:2014qua} to unity without loss of generality. If the Page curve \emph{does} turn over, it seems that it would be problematic to fix $c$ \cite{AP} if the evaporation is not adiabatic. However in our model, the entanglement entropy never decreases. See below.} \cite{Bianchi:2014qua}
\be S(u) = -\frac{1}{12} \ln p'(u), \ee 
and can also be expressed in terms of the mirror trajectory (by using $p'(u) = \frac{1+\dot{z}(t)}{1-\dot{z}(t)}$),
\be \label{entropy} S(t) = -\frac{1}{6} \tanh^{-1} [\dot{z}(t)], \ee
where $z(t)$ is the trajectory motion of the moving mirror, and the dot represents the time derivative.  In the case we consider below, the faster the mirror moves, the greater the entropy.   Unitary in this sense means the entropy must achieve a constant value in the far past and far future (not necessarily zero).  This underscores that eternally thermal radiation is not the end state.  That is, evolution from a possible initially pure state, to a final mixed state does not occur.  In this context, radiation of energy flux stops and one can acheive the initial pure state, preserving unitarity.  We will return to this issue in the next section.

The energy flux may also be expressed at the mirror in terms of the entropy as a function of time:
\be 2 \pi F(t) = e^{-12 S(t)} \cosh^2 [6 S(t)] ( 6 \dot{S}(t)^2 \tanh[6 S(t)] + \ddot{S}(t)), \ee
which is consistent with the entropy-energy relationship of Bianchi-Smerlak\cite{Bianchi:2014qua}
\be 2\pi F(u) = 6 S'(u)^2 + S''(u). \ee


\subsection{The Mirror, the Flux, the Energy and the Beta Coefficients}\label{sec:solution}

To determine whether time dependent particle production contains signatures of energy flux, we seek a moving mirror trajectory that has prominent or interesting emission character.  We have found such a dynamic solution.
A characteristic feature of this mirror, besides the NEF, is a central pulse of positive energy flux. The mirror emits this before finally emitting its NEF and dying back down to zero energy emission. 
We refer to such a pulse as the ``birth cry'' of the black hole. 

Of course, the features of the energy flux are completely trajectory-dependent; we do not claim that this model actually corresponds to any realistic evaporating black hole spacetime; for this work we only seek a simple model in which we can study whether energy flux (in particular, the NEF) is detectable via particle count. It is nevertheless worth pointing out that our dynamic solution is new --- we know of no unitary solution with a positive energy pulse over the thermal line, like the one exhibited here. 

\begin{figure}[ht]
\begin{center}
\mbox{\subfigure{\includegraphics[width=3.0in]{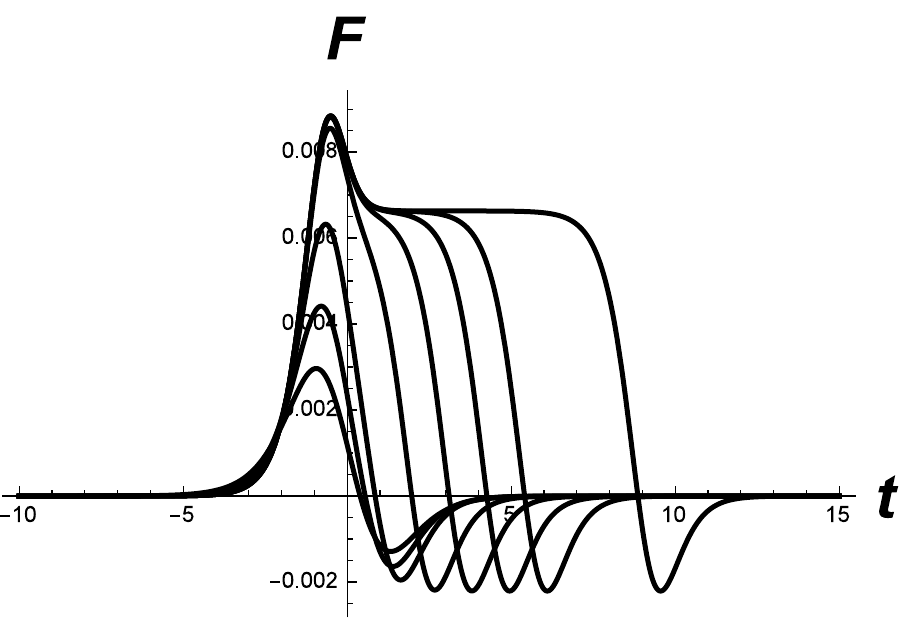}}\quad
\subfigure{\includegraphics[width=3.0in]{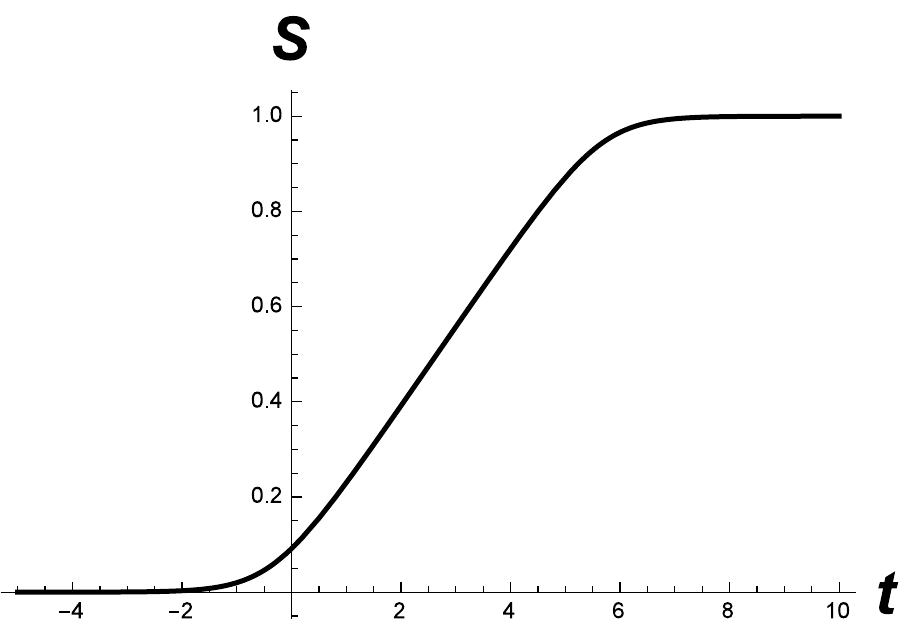} }}
\caption{\label{fig:flux} \textbf{Left:}  Successive plots of the energy flux, $F(t):=\la T_{uu} \ra$, Eq.(\ref{energyflux}), from smallest peak to the largest peak, with varying limiting mirror speeds, $\xi=0.7,0.8,0.9,0.99$ and $\xi=1-0.1^x$ where $x=3,4,5,8$.  Notice the three prominent flux characteristics: the positive energy hyper-thermal flare, the thermal plateau, and the NEF.  The positive energy hyper-thermal pulse and the thermal plateau occur only for a very fast final coasting speed.  The NEF is a more robust feature of the flux than either the thermal plateau or the hyper-thermal flare (though the local maximum on the curve is also robust).  The acceleration parameter is set to $\kp = 1$. \textbf{Right:} The von Neumann entanglement entropy as a function of time, with $\xi = 0.999987$ and $\kappa = 1$.  One sees that it approaches a constant value, underscoring the final asymptotic coasting speed of the mirror and the eventual end of Hawking radiation. There is no turn over of the Page curve.} 
\end{center}
\end{figure}

Starting off asymptotically static, this mirror has no asymptote, ultimately producing a finite total energy\footnote{This model therefore respects the observation that evaporating black holes should create finite energy flux\cite{kodama, hiscock1, hiscock2, kuroda}.}.  The motion of the mirror is given by the trajectory

\be \label{trajectory} z(t) = -\f{1}{2\kappa}\ln (e^{2\kp\xi t} + 1), \ee
where $0<\xi<1$ is the final speed of the mirror as $t\rightarrow\infty$, and $\kp>0$ is an acceleration parameter (not the acceleration, $\alpha(t) \neq \kappa$).  The proper acceleration, $\alpha = \gamma^3\ddot{z}$, is found to be, of course, non-uniform:
\be \label{propacc} \alpha(t) = -\frac{4 \kappa  \xi ^2 \text{sech}^2(\kappa  \xi  t)}{\left[-2 \xi ^2+\xi ^2 \left(\text{sech}^2(\kappa  \xi  t)-2 \tanh (\kappa  \xi  t)\right)+4\right]^{3/2}}. \ee
The negative sign on Eq.(\ref{propacc}) indicates the motion is to the left.  The acceleration has asymptotic behavior such that
\be \lim_{t\rightarrow \pm\infty} \alpha(t) = 0, \ee
making this trajectory asymptotically inertial.  However, although asymptotically inertial, and initially asymptotically static,
\be \lim_{t\rightarrow -\infty} \dot{z}(t) = 0, \ee
this mirror does not approach a future asymptotically static resting state because its future asymptotic velocity is
\be \lim_{t\rightarrow +\infty} \dot{z}(t) = \xi, \ee
making this trajectory \emph{asymptotically coasting}.  The drifting feature of this mirror means it is an exact model for a remnant \cite{aharonov}, as anticipated by Wilczek in \cite{Wilczek:1993jn}.  For a recent review of black hole remnants, see \cite{Chen:2014jwq}.  The entropy of this trajectory using Eq.(\ref{entropy}) is plotted in Figure (\ref{fig:flux}).  

In this particular remnant scenario, note that the entanglement entropy is monotonically increasing, i.e., there is no turn-over of the Page curve (this is not a necessary feature of a remnant; see \cite{Chen:2014jwq}). This means that the radiation is never actually purified. In the black hole evolution context, unitarity is preserved in the sense that the pure state remains pure if we consider both the exterior and the interior of the black hole remnant \cite{aharonov}. One may object that such a remnant scenario is problematic due to the well-known infinite production problem (among other objections), but this is outside the scope of the current work. Interested readers are referred to \cite{Chen:2014jwq} for detailed discussions. 
Note also that our remnant model is not in conflict with the argument for non-monotonic mass loss of Bianchi-Smerlak \cite{Bianchi:2014vea}, since the latter does not require that the entanglement entropy vanishes at the end stage of the evolution; as long as the entropy goes to a constant value, Bianchi-Smerlak's argument follows through. 

 The energy flux, in terms of any mirror trajectory, $z(t)$, was first derived by Davies and Fulling \cite{Davies:1976hi}. It is
\be\label{stressz} F(t):=\la T_{uu} \ra = \f{\dddot{z}(\dot{z}^2-1)-3\dot{z}\ddot{z}^2}{12\pi(\dot{z}-1)^4(\dot{z}+1)^2},\ee
where the dots are time derivatives. The energy flux, using the trajectory of Eq.(\ref{trajectory}), is therefore:
\be \label{energyflux}F(t) =  \frac{\kappa ^2 \xi ^3 e^{2 \kappa  \xi  t} \left(e^{2 \kappa  \xi  t}+1\right) \left[\left(\xi ^2-1\right) e^{6 \kappa  \xi  t}+\left(2 \xi ^2-1\right) e^{4 \kappa  \xi  t}+e^{2 \kappa  \xi  t}+1\right]}{3 \pi  \left[(\xi -1) e^{2 \kappa  \xi  t}-1\right]^2 \left[(\xi +1) e^{2 \kappa  \xi  t}+1\right]^4},\ee
and contains three major characteristics: a hyper-thermal positive pulse, a thermal plateau and NEF, see Figure (\ref{fig:flux}).  In order to investigate particle production, it proves useful and possible to write the beta Bogolubov coefficient in terms of the trajectory, $z(t)$  (For more details, see Good-Anderson-Evans \cite{Good:2013lca}):
\be \label{Bfromz} \beta_{\omega\omega'} = \frac{1}{4\pi \sqrt{\omega\omega'}}\int_{-\infty}^{\infty} dt\; \left[ e^{-i \omega_+ t + i \omega_- z(t)}(\omega_+ \dot{z}(t) - \omega_-)\right], \ee
where $\omega_{\pm} := \omega \pm \omega'$. This integral can be evaluated with the trajectory of Eq.(\ref{trajectory}).  The solution is in terms of an Euler integral of the first kind, i.e. the beta coefficient is a beta function:
\be \label{particlebeta} \beta_{\w'\w} =\frac{\sqrt{\w  \w'}}{2\pi  \kp  \xi  (\w -\w')} B\left[\frac{i (\omega +\w')}{2\kp  \xi }, -\frac{i ((1+\xi)\omega +(1-\xi)\w')}{2\kappa  \xi }\right]. \ee
The total energy emitted throughout the motion, emitted to an observer at $\mathscr{I}_R^+$ is
\be \label{totalenergy} E=\frac{\kappa}{48 \pi} \frac{ (\xi +1)^2 \tanh ^{-1}(\xi )-\xi  (2 \xi +1)}{\xi +1}, \ee
which we can confirm from
\be \label{quantasum} \int_0^\infty \omega \left[\int_0^\infty |\beta_{\omega\omega'}|^2 d\omega'\right] d\omega = \int_{-\infty}^{\infty} \la T_{uu}(t) \ra (1-\dot{z})dt. \ee 
This verifies that these results are consistent.  This also adds weight to the notion that the particles carry the energy, in the general sense that the total summation of the energies of each quanta is equal to the integral over the energy flux:
\be \int_0^\infty \omega \la N_\omega \ra d\omega = \int_{-\infty}^{\infty} F(u) du. \ee 

\section{The Energy Flux: A Thermal Plateau and Two Flares}

A period of thermal emission occurs at extremely high coasting speeds, giving a thermal plateau, which is, in the limit $\xi \rightarrow 1$, located at
\be \label{plateau} \lim_{t\rightarrow\infty} F(t) = \f{\kp^2}{48 \pi}. \ee
By construction, this is the same as the constant flux produced by the (eternally thermal) Carlitz-Willey trajectory \cite{Carlitz:1986nh}, which radiates a thermal Planckian distribution of particles for all times.  
A positive energy flare peaks, in the $\xi \rightarrow 1$ limit, at
\be F(t)^{\text{peak}} = \f{4}{3} \f{\kp^2}{48 \pi}, \ee
at time
\be t^{\text{peak}} = \f{1}{2\kp}\ln{\f{\sqrt{3}-1}{2}} \approx -\f{1}{2\kp}. \ee 
The energy flux that exceeds Eq.(\ref{plateau}) is above the thermal line and correspondingly we may also refer to it as a ``hyper-thermal'' flare.  
This is the black hole's birth cry in this model.
Even in the case of final coasting speeds where the thermal plateau is unpronounced in the energy flux, a hyper-thermal flare can exist.  For instance, when $\xi = 0.99$, the energy flux exceeds the value in Eq.(\ref{plateau}) but there is only a nascent formation of a thermal plateau. For low speeds of $\xi$, there is still a local maximum in the energy flux where the positive energy peaks but since this energy does not exceed Eq.(\ref{plateau}), we do not call it hyper-thermal --- the cry is a fainter one.

For $\xi$ very close to the speed of light ($\xi = 0.99999999 = 1-0.1^8$, no formal limit is taken), the energy flux has a minimal negative value  
\be F(t)^{\text{min}} = -\f{1}{3}\f{\kp^2}{48\pi}, \ee
which is at the bottom of the valley of the negative energy flux.

\section{The Particle Production}

Particle production has a different definition than energy flux \cite{Walker:1982, Walker:1984vj}.  However they are closely related by the trajectory of the mirror. The information encoded in the trajectory provides the link between the energy flux and the particle production.  It is already known that a mirror can radiate zero, or even negative, energy flux while still producing particles \cite{Walker:1984vj}.

The exact solution we considered has this interesting ``hyper-thermal'' flare, defined as that instantaneous energy flux which exceeds the thermal line at $\kappa^2/48\pi$.  It peaks before the thermal plateau at high final speeds making it a salient feature of the energy.  Does this behavior shows up in the particle creation as well?  We find indeed it does. The physical interpretation is that, not only do the particles emitted carry the total energy radiated in such a unitary situation but that they carry signatures of the instantaneous energy flux emission.  This is a stronger evidence than that of the consistency agreement of the bare total energy emission from the summation of quanta energy of Eq.(\ref{quantasum}) alone.  

\begin{figure}[!h]
\centering
\mbox{\subfigure{\includegraphics[width=3.0in]{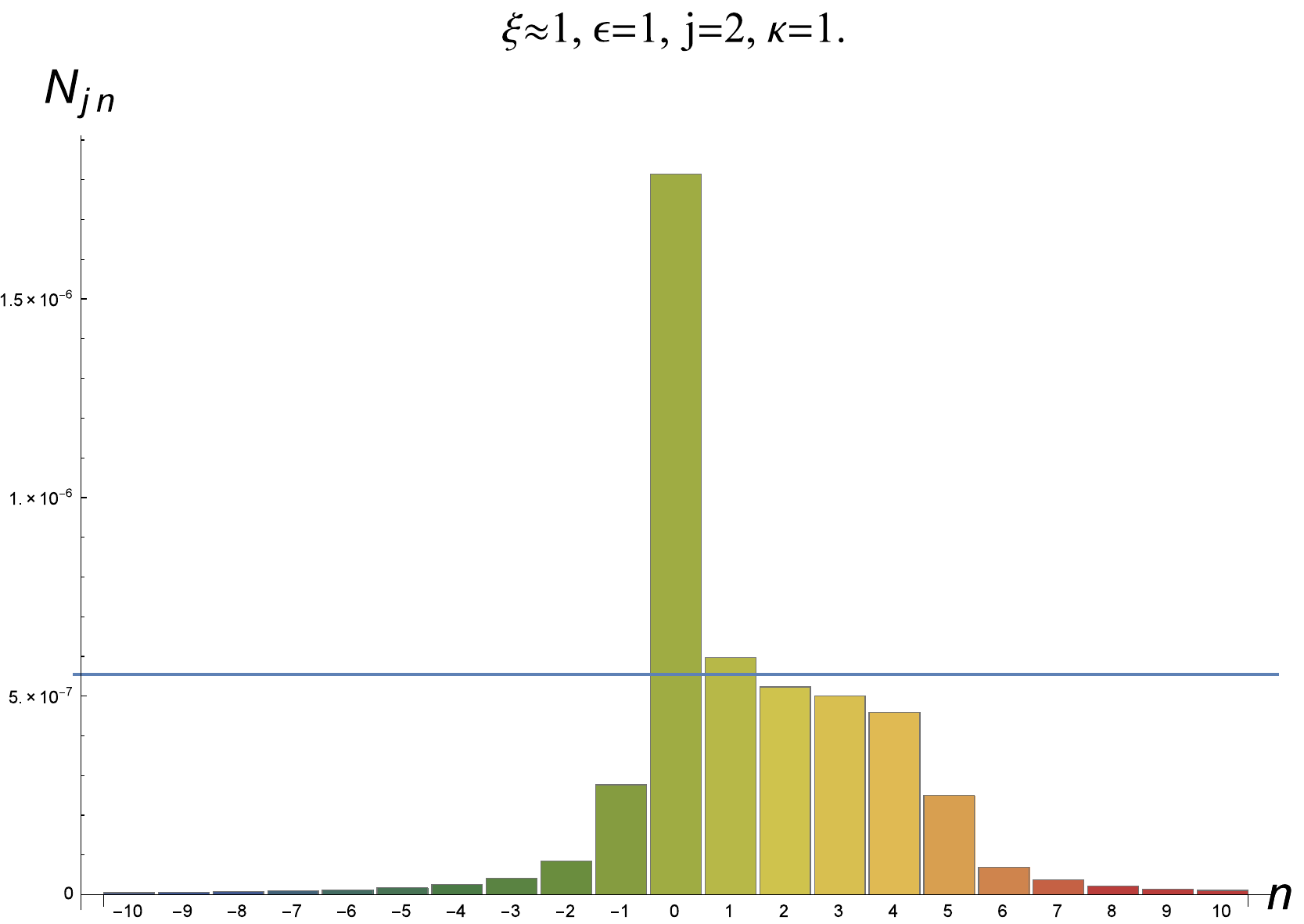}}\quad
\subfigure{\includegraphics[width=3.0in]{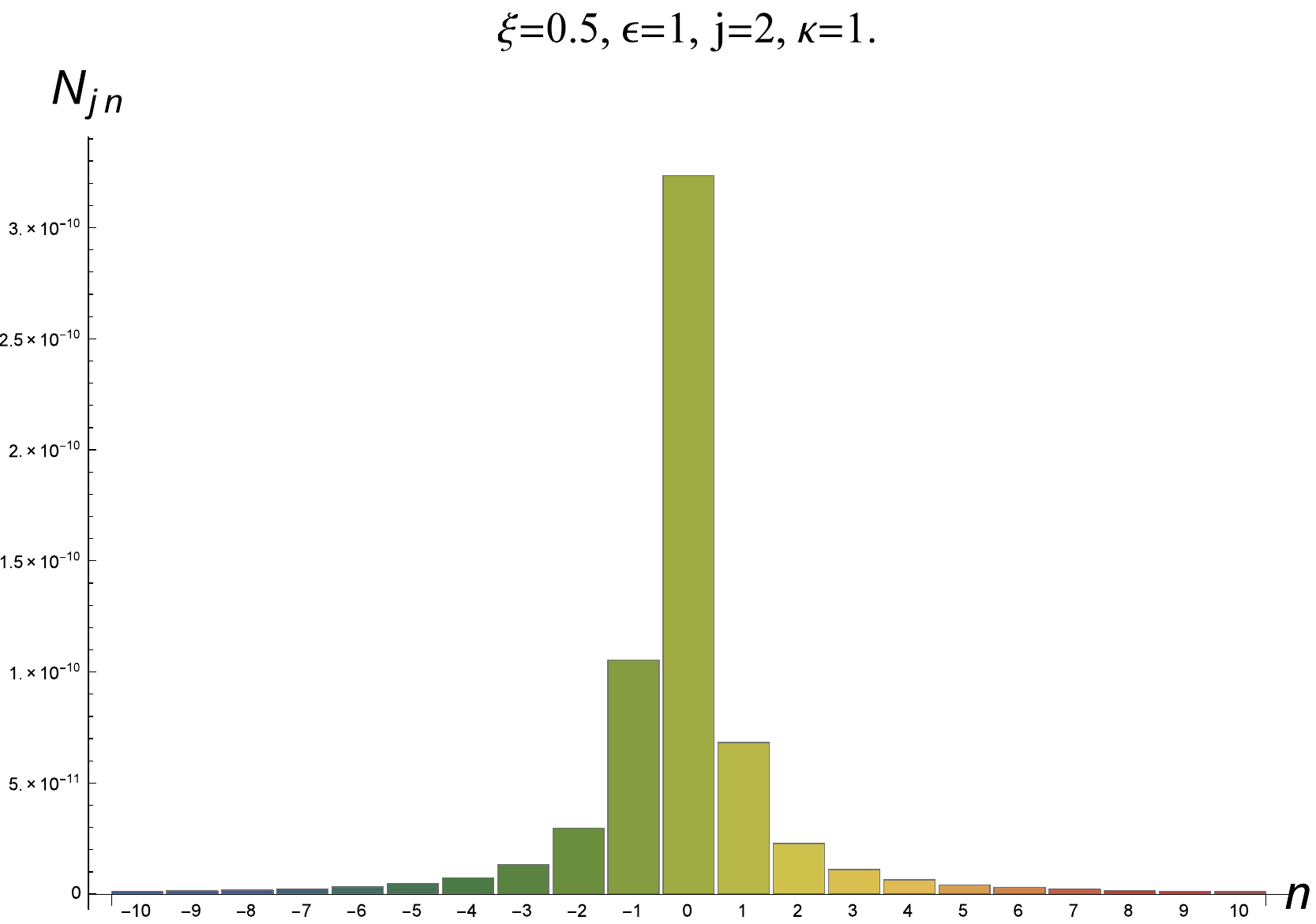} }}
\caption{\textbf{Left:} The particle production with final drifting speed very close to the speed of light, $\xi = 0.999999999999999 = 1-0.1^{15}$.  The acceleration parameter is set to $\kp = 1$.  Notice the hyper-thermal pulse of particles where the production exceeds the horizontal constant emission line. The horizontal line is the exact solution to particle production with constant flux thermal emission \cite{Good:2013lca}. \textbf{Right:} The particle production with final drifting speed at half the speed of light, $\xi = 0.5$.  The acceleration parameter is also set to $\kp =1$.   In order to compare with the left graph which has a much faster drifting speed, notice that at half the speed of light there is a more extreme drop off of particle production in time, i.e. increasing the final coasting speed produces a thermal plateau.  \label{fig:ppfast}}
\end{figure}

See Figure (\ref{fig:ppfast}), in which we constructed ``packetized'' beta Bogolubov coefficients
\be
\beta_{jn\w'} =
\f{1}{\sqrt{\epsilon}}\int_{j\epsilon}^{(j+1)\epsilon}
d\w \; \left[e^{\frac{2\pi i \w n}{\epsilon}} \beta_{\w\w'} \right]\;,
\label{beta-packet}
\ee
using the orthonormal and complete wave packets of Hawking \cite{Hawking2} and calculated the particle count,
\bea  
\la N_{jn}\ra &=& \int_0^\infty d\w' |\beta_{jn,\w'}|^2   \;,  
\nonumber 
\\
&=& \int_0^\infty d\w' \int_{j \epsilon}^{(j+1)\epsilon} 
\frac{d \w_1}{\sqrt{\epsilon}} \int_{j \epsilon}^{(j+1)\epsilon} 
\frac{d \w_2}{\sqrt{\epsilon}} \left[ e^{\frac{2 \pi i(\w_1- \w_2)n}{\epsilon}}
\beta_{\w_1 \w'} \beta^{*}_{\w_2 \, \w'} \right] \;. 
\label{Njn} 
\eea
Particles that arrive at $\mathscr{I}^+_R$ can be investigated by obtaining these wave packet coefficients straightforwardly from the 
coefficients $\beta_{\w\w'}$ of Eq.(\ref{particlebeta}). With the Bogolubov coefficients for the packets in hand, the average number 
of particles produced for given values of $n$ and $j$ is determined.  They arrive at $\mathscr{I}^+_R$ in the range of frequencies $j \epsilon \leqslant \omega \leqslant (j+1) \epsilon$ and in the range of times $ (2\pi n - \pi)/\epsilon \leqslant u \leqslant (2 \pi n + \pi)/\epsilon$.  Further details on how these packets are constructed from the modes themselves or for non-asymptotically inertial trajectories can be found in \cite{Good:2013lca}.

\section{Conclusion: Birth Cry vs. Death Gasp}

In this work we have shown that particle production and energy flux, evolved in time, exhibit manifest signatures corresponding to the instantaneous radiation.  
In particular, 
we describe a new, explicit, moving mirror solution which exhibits
the thermal plateau, as well as the positive energy burst (the black hole's ``birth cry'') before it. Both of these features can be detected by asymptotic observers with particle detectors. 
This moving mirror trajectory also gives rise to negative energy flux (NEF), and by the result of Bianchi-Smerlak \cite{Bianchi:2014vea,Bianchi:2014qua}, also to the non-monotonic mass loss of the corresponding black hole.  The question is then, whether one could also detect such flux from particle count at infinity.

We find no signature of negative energy flux in the particle emission.  We do not perceive a black hole's dying gasp via particle emission, in stark contrast to its birth cry.  The lack of signatures of NEF in the particle production is surprising in light of the evidence that the other energy flux characteristics have explicit signatures in the particle production.  One might expect a rapid decrease or increase in particle emission due to the NEF, i.e. a ``second flare''; or one might also anticipate zero particle production at an instant in time that corresponds to the zero energy flux emission point.  We have not found these anticipations to have materialized.  It should be obvious that what one does not expect is negative particle production because the Bogolubov coefficients are complex conjugated,
\be \la N_\omega \ra = \int_0^\infty |\beta_{\omega\omega'}|^2 d\omega', \ee
leaving the computed occupation number always positive.  Still, the NEF is received at infinity accompanied by real particle production but the character of the NEF is ``hidden''. Presumably, the NEF is carried by the emitted particles, however the particle emission in time gives no hint of NEF (or, correspondingly, an increase in the mass of the black hole).  This detector, which bleeps, upon a detection, reveals no suspicious patterns that the black hole is emitting NEF or swelling up, since the particle production, after its central peak, monotonically decreases.  Absence of evidence is not evidence of absence, however, and therefore we cannot rule out the possibility that the signatures of NEF may still be exhibited in the particle production.  Evidence might emerge readily in mirror trajectories which have salient features of energy flux and achieve asymptotically static motion.  However, it does seem likely that one could not detect negative energy flux by simply looking at particle production alone; instead one may need to use some other probes.

It is also worth noting that in a recent work of Abdolrahimi-Page \cite{AP}, the negative energy flux in the case of an asymptotically flat Schwarzschild black hole is shown to be extremely tiny. They found that the mass increase of the black hole is less than 0.09\% of the energy of a \emph{single} quantum of the energy of the Hawking temperature of said black hole at that time, and is therefore unlikely to be detectable, considering that such a signature would in addition be swamped by the noise of quantum fluctuations. However, note that in our model the aim is purely to study the detectability of negative energy flux at infinity via particle production evolution signatures, regardless of whether the negative energy flux is tiny or not.

\section*{Acknowledgment}
MRRG thanks NORDITA, where part of this work was completed, for its hospitality. 
YCO thanks the Yukawa Institute for Theoretical Physics for hospitality during the molecule-type workshop on ``Black Hole Information Loss Paradox''; he also thanks Don Page for fruitful discussion during the workshop.



\begin{thebibliography}{99}

\bibitem{Hawking2}
S. W. Hawking, ``Particle Creation by Black Holes'', Commun. Math. Phys. \textbf{43} (1975) 199.


\bibitem{page1}
D. N. Page, ``Average Entropy of a Subsystem'', Phys. Rev. Lett. \textbf{71} (1993) 1291, \href{http://arxiv.org/abs/gr-qc/9305007}{[arXiv:gr-qc/9305007]}.

\bibitem{page1b}
D. N. Page, ``Information in Black Hole Radiation'', Phys. Rev. Lett. \textbf{71} (1993) 3743, \href{http://arxiv.org/abs/hep-th/9306083}{[arXiv:hep-th/9306083]}.

\bibitem{page2}
D. N. Page, ``Time Dependence of Hawking Radiation Entropy'', JCAP \textbf{1309} (2013) 028 , \href{http://arxiv.org/abs/1301.4995}{[arXiv:1301.4995 [hep-th]]}.

\bibitem{sabine}
S. Hossenfelder, ``Disentangling the Black Hole Vacuum'', 	Phys. Rev. D \textbf{91} (2015) 044015, \href{http://arxiv.org/abs/1401.0288}{[arXiv:1401.0288 [hep-th]]}.

\bibitem{1409.7754}
M. Visser, ``Thermality of the Hawking Flux'', \href{http://arxiv.org/abs/1409.7754}{[arXiv:1409.7754 [gr-qc]]}.

\bibitem{Bianchi:2014vea} 
  E.~Bianchi, M.~Smerlak,
  ``Last Gasp of a Black Hole: Unitary Evaporation Implies Non-Monotonic Mass Loss'',
  Gen.\ Rel.\ Grav.\  {\bf 46} (2014) 1809,
  \href{http://arxiv.org/abs/1405.5235}{[arXiv:1405.5235 [gr-qc]]}.
	
\bibitem{Bianchi:2014qua} 
  E.~Bianchi, M.~Smerlak,
  ``Entanglement Entropy and Negative Energy in Two Dimensions'',
  Phys.\ Rev.\ D {\bf 90} (2014) 041904,
  \href{http://arxiv.org/abs/1404.0602}{[arXiv:1404.0602 [gr-qc]]}.
	
\bibitem{HMF}
M. Hotta, J. Matsumoto, K. Funo, ``Black Hole Firewalls Require Huge Energy of Measurement'', Phys. Rev. D \textbf{89} (2014) 12, 124023, \href{http://arxiv.org/abs/1306.5057v4}{[arXiv:1306.5057 [quant-ph]]}.

\bibitem{amps}
A. Almheiri, D. Marolf, J. Polchinski, J. Sully, ``Black Holes: Complementarity or Firewalls?'', JHEP \textbf{1302} (2013) 062, \href{http://arxiv.org/abs/1207.3123}{[arXiv:1207.3123 [hep-th]]}.

\bibitem{apologia}
A. Almheiri, D. Marolf, J. Polchinski, D. Stanford, J. Sully,
``An Apologia for Firewalls'', JHEP \textbf{1309} (2013) 018, \href{http://arxiv.org/abs/1304.6483}{[arXiv:1304.6483 [hep-th]]}.


\bibitem{sam}
S. L. Braunstein, S. Pirandola, K. \.Zyczkowski, ``Better Late than Never: Information Retrieval from Black Holes'', Phys. Rev. Lett. \textbf{110} (2013) 101301, \href{http://arxiv.org/abs/0907.1190}{[arXiv:0907.1190 [quant-ph]]}.
	
\bibitem{weinstein}
M. Weinstein, ``Moving Mirrors, Black Holes, Hawking Radiation and All That'', Nucl. Phys. Proc. Suppl. \textbf{108} (2002) 68, \href{http://arxiv.org/abs/gr-qc/0111027}{[arXiv:gr-qc/0111027]}.
	
\bibitem{Wilson:2011} 
  C.~M.~Wilson, G.~Johansson, A.~Pourkabirian, J.~R.~Johansson, T.~Duty, F.~Nori, P.~Delsing,
  ``Observation of the Dynamical Casimir Effect in a Superconducting Circuit,'' Nature \textbf{479} (2011) 376.
  
\bibitem{Rego:2014wta} 
  A.~L.~C.~Rego, C.~Farina, H.~O.~Silva, D.~T.~Alves, ``New Signatures of the Dynamical Casimir Effect in a Superconducting Circuit,''
  Phys.\ Rev.\ D {\bf 90} (2014) 025003 
  \href{http://arxiv.org/abs/1405.3720}{[arXiv:1405.3720 [quant-ph]]}.
	
\bibitem{Davies:1976hi} 
  P.~C.~W.~Davies, S.~A.~Fulling,
  ``Radiation from a Moving Mirror in Two-Dimensional Space-Time Conformal Anomaly,''
  Proc.\ Roy.\ Soc.\ Lond.\ A {\bf 348} (1976) 393.
	
\bibitem{Davies:1977yv} 
  P.~C.~W.~Davies, S.~A.~Fulling,
  ``Radiation from Moving Mirrors and from Black Holes,''
  Proc.\ Roy.\ Soc.\ Lond.\ A {\bf 356}  (1977) 237.
	
	
\bibitem{Carlitz:1986nh} 
  R.~D.~Carlitz, R.~S.~Willey,
  ``Reflections On Moving Mirrors,''
  Phys.\ Rev.\ D {\bf 36} (1987) 2327.
	
	
\bibitem{Walker:1982} 
  W.~R.~Walker, P.~C.~W.~Davies,
  ``An Exactly Soluble Moving-Mirror Problem'',
  J. Phys.\ A: Math. Gen. {\bf 15} (1982) L477.
	
\bibitem{Good:2013lca} 
  M.~R.~R.~Good, P.~R.~Anderson, C.~R.~Evans,
  ``Time Dependence of Particle Creation from Accelerating Mirrors,''
  Phys.\ Rev.\ D {\bf 88} (2013) 025023,
  \href{http://arxiv.org/abs/1303.6756}{[arXiv:1303.6756 [gr-qc]]}.
	
	\bibitem{AP}
	S. Abdolrahimi, D. N. Page, ``Hawking Radiation Energy and Entropy from a Bianchi-Smerlak Semiclassical Black Hole'', \href{http://arxiv.org/abs/1506.01018}{[arXiv:1506.01018 [hep-th]]}.
	
	\bibitem{kodama}
	H. Kodama, ``Conserved Energy Flux for the Spherically Symmetric System and the Backreaction Problem in the Black Hole Evaporation '', Prog. Theor. Phys.\textbf{63} (1980) 1217. 
	
	\bibitem{hiscock1}
	W. A. Hiscock, ``Models of Evaporating Black Holes. I'', Phys. Rev. D \textbf{23} (1981) 2813.
	
	\bibitem{hiscock2}
	W. A. Hiscock, ``Models of Evaporating Black Holes. II. Effects of the Outgoing Created Radiation'', Phys. Rev. D \textbf{23} (1981) 2823.
	
	\bibitem{kuroda}
	Y. Kuroda, ``Vaidya Spacetime as an Evaporating Black Hole'', Prog. Theor. Phys. \textbf{71} (1984) 1422.
	
	\bibitem{aharonov}
	Y. Aharonov, A. Casher, S. Nussinov, ``The Unitarity Puzzle and Planck Mass Stable Particles'', Phys. Lett. B \textbf{191} (1987) 51.

\bibitem{Wilczek:1993jn} 
  F.~Wilczek,
  ``Quantum Purity at a Small Price: Easing a Black Hole Paradox'', 
  In *Houston 1992, Proceedings, Black Holes, Membranes, Wormholes and Superstrings* 1-21, and Inst. Adv. Stud. Princeton - IASSNS-HEP-93-012 (93/02,rec.Mar.) 19 p. (306377),
 \href{http://arxiv.org/abs/hep-th/9302096}{[hep-th/9302096]}.
	
\bibitem{Chen:2014jwq} 
  P.~Chen, Y.~C.~Ong, D.-h.~Yeom,
  ``Black Hole Remnants and the Information Loss Paradox'',
  \href{http://arxiv.org/abs/1412.8366}{[arXiv:1412.8366 [gr-qc]]}.
	
\bibitem{Walker:1984vj} 
  W.~R.~Walker,
  ``Particle and Energy Creation by Moving Mirrors'',
  Phys.\ Rev.\ D {\bf 31} (1985) 767.
	
	

	
	


\end{thebibliography}
\end{document}